\documentclass[aps,prd,superscriptaddress,nofootinbib,amsmath,amsfonts,preprintnumbers,groupedaddress,showpacs,10pt,english]{revtex4-1}
\usepackage{amsmath}
\usepackage{amssymb}
\usepackage{babel}
\usepackage{wrapfig}
\usepackage{cancel}

\usepackage{relsize,exscale}
\usepackage{enumerate}
\makeatletter

\usepackage{array,multirow,graphicx}
\usepackage{dcolumn}
\usepackage{newlfont}
\usepackage{bm}
\usepackage[colorlinks,citecolor=blue,urlcolor=blue,linkcolor=blue]{hyperref}
\usepackage[figtopcap]{subfigure}
\usepackage{color}
\usepackage{verbatim}

\def\be{\begin{align}}
\def\ee{\end{align}}
\def\bea{\begin{eqnarray}}
\def\eea{\end{eqnarray}}
\def\bal{\begin{align}}
\def\eal{\end{align}}

\newcommand{\arctanh}{{arctanh}}

\usepackage{scalerel}
\usepackage{tikz}
\usetikzlibrary{svg.path}
\definecolor{orcidlogocol}{HTML}{A6CE39}
\tikzset{
 orcidlogo/.pic={
 \fill[orcidlogocol] svg{M256,128c0,70.7-57.3,128-128,128C57.3,256,0,198.7,0,128C0,57.3,57.3,0,128,0C198.7,0,256,57.3,256,128z};
 \fill[white] svg{M86.3,186.2H70.9V79.1h15.4v48.4V186.2z}
 svg{M108.9,79.1h41.6c39.6,0,57,28.3,57,53.6c0,27.5-21.5,53.6-56.8,53.6h-41.8V79.1z M124.3,172.4h24.5c34.9,0,42.9-26.5,42.9-39.7c0-21.5-13.7-39.7-43.7-39.7h-23.7V172.4z}
 svg{M88.7,56.8c0,5.5-4.5,10.1-10.1,10.1c-5.6,0-10.1-4.6-10.1-10.1c0-5.6,4.5-10.1,10.1-10.1C84.2,46.7,88.7,51.3,88.7,56.8z};}}
\newcommand\orcid[1]{\href{https://orcid.org/#1}{\mbox{\scalerel*{
\begin{tikzpicture}[yscale=-1,transform shape]
\pic{orcidlogo};
\end{tikzpicture}
}{|}}}}
\graphicspath{{./}{Figs/}}
\begin{document}
\date{\today}
\title{Stellar isotropic model in the symmetric teleparallel equivalent of general relativity theory}

\author{G.~G.~L.~Nashed$^{1}$\orcid{0000-0001-5544-1119}}\email{nashed@bue.edu.eg}
\author{Amare Abebe$^{2,3}$}\email{Amare.Abebe@nithecs.ac.za}
\affiliation{$^{1}$ Centre for Theoretical Physics, The British University in Egypt, P.O. Box
43, El Sherouk City, Cairo 11837, Egypt \\
$^{2}$ Centre for Space Research, North-West University, Potchefstroom, South Africa \\
$^{3}$ National Institute for Theoretical and Computational Sciences (NITheCS), South Africa}

\begin{abstract}
Recently, the theory of symmetric teleparallel equivalent of general relativity (STEGR) has gained much interest in the cosmology and astrophysics community. Within this theory, we discuss the method of deriving a stellar isotropic model. In this respect, we implement the equations of motion of STEGR theory to a spacetime that is symmetric in a spherical manner, resulting in a set of nonlinear differential equations with more unknowns than equations.  To solve this issue, we assume a special form of $g_{tt}$, and suppose a null value of the anisotropy to obtain the form of $g_{rr}$. We then investigate the possibility of obtaining an isotropic stellar model consistent with observational data.  To test the stability of our model, we apply the adiabatic index and the Tolman-Oppenheimer-Volkoff equation. Furthermore, we examine our model using different observed values of radii and masses of pulsars, showing that all of them fit in a consistent way.
\end{abstract}

\maketitle
\section{Introduction}\label{S1}

Modifications to Einstein's general relativity (GR) have gained significant interest for investigating the late-time accelerated expansion of the universe without the need to introduce dark energy. These modified theories have also produced promising results in relativistic cosmology, addressing many unresolved issues such as matter density fluctuations, the formation of large-scale structures, and inhomogeneities and anisotropies.  Recent observational data, as presented in \cite{perlmutter1997measurements, perlmutter1998discovery, perlmutter1999measurements, riess1998observational, Reichardt:2011yv}, indicate that our current universe is undergoing accelerated expansion. This provides clear evidence that the characteristics of gravity extend beyond GR. The fundamental mystery of cosmic expansion lies in the fact that most of our universe consists of material known as dark energy \cite{Barris:2003dq}.  The gravitational action can be described in various forms, such as  $f(T)$  where $T$ is the torsion scalar \cite{Cai:2015emx,Nashed:2015pga}, $f(R)$ where $R$  is the Ricci scalar \cite{delaCruz-Dombriz:2006kob}, or a combination of these, such as $f(R,T)$  \cite{Myrzakulov:2012qp,Nashed:2022zyi}, or $f(G)$ where $G$ is the Gauss-Bonnet invariant \cite{Betz:2008wy,Nashed:2018qag}, among others.

Recently, an additional modification to GR known as $f(Q)$  theory has been developed \cite{Capozziello:2022wgl}. This theory extends the symmetric teleparallel equivalent of GR (STEGR), characterized by a connection where both curvature and torsion vanish. Thus, STEGR is defined by a nonmetricity scalar field, $Q$, which describes geometric constructions equivalent to GR \cite{Khyllep:2021pcu}. Therefore, $f(Q)$ gravitational theory has garnered significant interest to explain various topics within the framework of cosmology \cite{DAgostino:2022tdk}. A detailed investigation of STEGR in the context of the background evolution of the universe has been conducted in \cite{Ayuso:2020dcu}. Various self-accelerating models are discussed in \cite{Junior:2023qaq}, which analyze the cosmological expansion history in $f(Q)$ using observations such as high-redshift Hubble diagrams from SNIa, baryon acoustic oscillations (BAO) and the CMB shift factor. All these observations, based on different distance measurements, are sensitive solely to the expansion history.

Recent observations by events such as GW170817 and GW190814, in collaboration with LIGO-VIRGO, have motivated researchers to increase their focus on modeling compact objects involved in binary mergers, which act as gravitational wave sources. More precisely, the detection of gravitational waves from event GW190814 indicates that the signals came from the combination of a black hole weighing between 22.2 and 24.3 solar masses and a compact object ranging  between 2.50 and 2.67 solar masses. On the other hand, the GW170817 phenomenon is linked to the combination of two neutron stars with sizes between 0.86 and 2.26 solar masses. Theorists have looked into modified gravity theories to understand how stellar masses greater than 2 solar masses can exist within the framework of standard GR without the need for exotic matter distributions or rotation.
  Among the most widely explored modifications in recent years is the $f(Q)$ gravity, and this model has shown significant predictive capabilities in both  cosmological and astrophysical contexts. A metricity factor with a power-law form, given by $f(Q)= a+bQ^n$  was discussed in  \cite{Capozziello:2022wgl}.
An anisotropic hybrid stars have been constructed  using a singularity-free gravitational potentials method, according to the Tolman-Kuchowicz   ansatz. In this model, within the context of $f(Q)$ gravity, the stellar fluid is made up of a mix of strange quark matter (SQM) and regular baryonic matter (OBM) \citep{Bhar:2023yrf}. Moreover, Bhar et al. added the MIT bag model equation of state (EOS) to enhance the gravitational aspects of the model.  Their models were free of singularities and encompassed a range of stellar masses, including compact objects within the mass range required for the secondary component of the GW 190814 event \citep{Bhar:2024vxk}. The mass-gap dilemma in gravitational events has posed several challenges for researchers in recent years. The latest addition to this category is a pulsar with a mass ranging from 2.09 to 2.71 solar masses, identified as a part of a binary system detected during the MeerKat survey \citep{Barr:2024wwl}.

Within this study we are going to use $f(Q)$ gravitational theory because it possesses (i) a long-standing historical context known as ``teleparallelism" or ``teleparallel gravity'' which was put forth by Einstein \cite{inbook} in 1928 in relation to his GR, and (ii) it has recently garnered attention from scientists \cite{Pereira:2013qza,Bahamonde:2021gfp} in both cosmological \cite{Dialektopoulos:2019mtr,BeltranJimenez:2019tme,Barros:2020bgg,Bajardi:2020fxh,Anagnostopoulos:2021ydo} and astrophysical \cite{Mandal:2020lyq,Flathmann:2020zyj,Hassan:2021egb,Banerjee:2021mqk,Chanda:2022cod,Maurya:2022vsn,Maurya:2022wwa,Errehymy:2022gws,Sokoliuk:2022bwi,Wang:2021zaz,Mustafa:2021bfs,Hassan:2022jgn,Hassan:2022hcb,Jan:2023djj,
Godani:2023nep,Mishra:2023bfe,Ditta:2023xhx,Maurya:2023szc,Junior:2023qaq,LHCb:2022bbb,Javed:2023vmb} context of research. Many scientists have studied wormhole geometries within $f(Q)$ gravity across various physical scenarios  \cite{Banerjee:2021mqk,Hassan:2021egb,Mustafa:2021bfs,Parsaei:2022wnu,Hassan:2022jgn,Hassan:2022hcb,Hassan:2022ibc,Jan:2023djj,Godani:2023nep,Mishra:2023bfe}. The purpose of this study is to derive an isotropic stellar model using STEGR gravitational theory.

The structure of the current study is as follows:  In Section \ref{S2} we present the basic structure of $f(Q)$ gravitational theory. In Section \ref{S3}, we utilize $f(Q)$ field equations for a spherically symmetric object with a matter source with anisotropy. The system generates three differential equations involving five unknown functions: energy-density, radial pressure, tangential pressure, and two metric potentials. As a result, two more restrictions were inserted.  We assume a form for $g_{tt}$, resembling the usual pattern found for the non-vacuum solutions. Moreover, we assume the vanishing of anisotropy to determine $g_{rr}$.  By gaining knowledge of the metric potentials $g_{tt}$ and $g_{rr}$, we can determine the mathematical formulas for pressure and density that satisfy the STEGR field equations.  In Section \ref{S4}, we describe the essential requirements in order for a stellar model to be considered a true compact star. In Section \ref{S5} We examine the importance of the obtained solution in light of the scenarios mentioned  in Section \ref{S4}. In Section \ref{S6}, we incorporate the Schwarzschild exterior solution into our model to fix the parameters of the model.
 The stellar $\textit{Cen X-3}$ is employed to calculate the parameters of the model, with a predicted mass of about $M= 1.49 \pm 0.49\, M_\odot$ and a diameter of roughly $R= 9.178 \pm 0.13$ km.  In Section \ref{S7}, we analyze the model's stability by studying the TOV equation as well as the adiabatic index.  In \ref{Conc}, a summary of our findings is presented.
\section{$f(Q)$ gravity}\label{S2}
The steps for the adapted $f(Q)$ gravity method are outlined in \citep{Zhao:2021zab,BeltranJimenez:2017tkd}: the functional form of $f(Q)$-gravity with matter sources is determined by
\begin{align}
\label{eq1}
\mathcal{S}=\int\frac{1}{2}\,f(Q)\sqrt{-g}\,d^4x+\int \mathcal{L}_m\,\sqrt{-g}\,d^4x\;.
\end{align}
The Lagrangian density for matter distribution is denoted by $\mathcal{L}_m$, $Q$ stands for the non-metricity scalar, and $f=f(Q)$ is a general function of $Q$. The energy-momentum tensor $T_{\alpha\nu}$ connected to Lagrangian $\mathcal{L}_m$ is expressed as:
\begin{align}
\label{eq2}
&& \frac{2}{\sqrt{-g}}\frac{\delta\left(\sqrt{-g}\,\mathcal{L}_m\right)}{\delta g^{\alpha\nu}}={T}_{\alpha\nu}\;.
\end{align}
The tensor $Q_{\lambda\alpha\nu}$ for the nonmetricity term is calculated as
\begin{align}\label{eq4}
&Q_{\lambda\alpha\nu}=\bigtriangledown_{\lambda} g_{\alpha\nu}=\partial_\lambda g_{\alpha\nu}-\Gamma^\delta_{\,\,\,\lambda \alpha}g_{\delta \nu}-\Gamma^\delta_{\,\,\,\lambda \nu}g_{\alpha \delta}\,, \, \, \mbox{where $\Gamma^\delta_{\,\,\,\alpha\nu}$ is the affine connection, defined as}\\
& \label{eq5}\Gamma^\delta_{\,\,\,\alpha\nu}=K^\delta_{\,\,\,\alpha\nu}+ L^\delta_{\,\,\,\alpha\nu}+\lbrace^\delta_{\,\,\,\alpha\nu} \rbrace, \quad \mbox{where} \quad  \frac{1}{2} T^\delta_{\,\,\,\alpha\nu}+T_{(\alpha\,\,\,\,\,\,\nu)}^{\,\,\,\,\,\,\delta}=K^\delta_{\,\,\,\alpha\nu}\;,\\
&\label{eq6} \frac{1}{2}Q^\delta_{\,\,\,\alpha\nu}-Q_{(\alpha\,\,\,\,\,\,\nu)}^{\,\,\,\,\,\,\delta}=L^\delta_{\,\,\,\alpha\nu},\quad  \frac{1}{2}g^{\delta\sigma}\left(\partial_\alpha g_{\sigma\nu}+\partial_\nu g_{\sigma\alpha}-\partial_\sigma g_{\alpha\nu}\right)=\lbrace^\delta_{\,\,\,\alpha\nu} \rbrace,
\end{align}
where  $\lbrace^\delta_{\,\,\,\alpha\nu} \rbrace$, $T^\delta_{\,\,\,\alpha\nu}$,  $L^\delta_{\,\,\,\alpha\nu}$, and $K^\delta_{\,\,\,\alpha\nu}$ are the Levi-Civita connection, torsion tensor, disformation, and contortion tensors are identified. Moreover, the antisymmetric component of the affine connection can be established as $T^\delta_{\,\,\,\alpha\nu}=2\Gamma^\lambda_{\,\,\,[\alpha\nu]}$. {
Eventually, the non-metricity scalar equation can be formulated as:
\begin{equation}\label{eq9}
Q=-Q_{\alpha\alpha\nu}\,P^{\alpha\alpha\nu}.
\end{equation}
The term $P^{\alpha\alpha\nu}$ offers a non-metricity counterpart. The tensor that corresponds is defined as:
\begin{equation}\label{eq7}
P^\alpha_{\,\,\mu\nu}=\frac{1}{4}\left[-Q^\alpha_{\,\,\mu\nu}+2 Q_{(\mu\,\nu)}^\alpha+Q^\alpha g_{\mu\nu}-\tilde{Q}^\alpha g_{\mu\nu}-\delta^\alpha_{(\mu}Q_{\nu)}\right].
\end{equation}
Here, $\tilde{Q}_\alpha$ and $Q_{\alpha}$ are two independent traces defined as follows:
\begin{equation}\label{eq8}
\tilde{Q}_\alpha=Q^\mu_{\,\,\,\,\alpha\mu}, \;\;\;Q_{\alpha}\equiv Q_{\alpha\,\,\,\mu}^{\,\,\mu}.
\end{equation}
In order to obtain the correct field equations for $f(Q)$-gravity, one must take the variation of the action (\ref{eq1}) w.r.t. the metric tensor $g^{\mu\nu}$. Therefore, the $f(Q)$ gravity field equations can be derived in the following manner:
\begin{align}
\label{eq10}
&& \hspace{.01cm}T_{\mu\nu}=\frac{2}{\sqrt{-g}}\bigtriangledown_\gamma\left(\sqrt{-g}\,f_Q\,P^\gamma_{\,\,\,\,\mu\nu}\right)+\frac{1}{2}g_{\mu\nu}f
+f_Q\big(P_{\mu\gamma \alpha}\,Q_\nu^{\,\,\,\gamma \alpha}-2\,Q_{\gamma \alpha \mu}\,P^{\gamma \alpha}_{\,\,\,\,\,\,\nu}\big),~~
\end{align}
where $f_Q=\frac{d f}{d Q}$. In this study, we are going to specialize to the simplest case of $f(Q)=Q$.


\section{spherically symmetric interior solution  }\label{S3}
Supposing that the spherical symmetric spacetime takes the shape:
 \begin{align} \label{met12}
&   ds^2=F^2(r)dt^2-\frac{1}{G(r)}\,dr^2-r^2(d\theta^2-\sin^2\theta d\phi^2)\,.  \end{align}
Here $F(r)$ and $G(r)$ are functions that are not identified. By utilizing the Eq.~(\ref{met12}), we obtain the Ricci scalar expressed as:
  \begin{align} \label{Ricci}
  { Q}(r)=-\frac {2GF'-FG'}{ {r} F}\,,
  \end{align}
  where $F\equiv F(r)$,  $G\equiv G(r)$,  $F'=\frac{dF}{dr}$, $F''=\frac{d^2F}{dr^2}$ and $G'=\frac{dG}{dr}$.
  Plugging Eq.
Eq. (\ref{met12}) with (\ref{Ricci}) in Eq. (\ref{eq10}) we get:
 \begin{align}\label{feqq}
&  \textrm{ The tt of Eq.~(\ref{eq10}) is expressed as:} \quad \epsilon=\frac{rG'+G-1}{r^2}\,,\quad \textrm{ The rr of Eq.~(\ref{eq10}) is expressed as:} \quad p=\frac{2rGF'+F(G-1)}{r^2F}\,,\nonumber\\
&  \textrm{ The}\, \mathrm{\theta\, \theta = \phi \, \phi}\textrm{ component of the  equation of motion (\ref{eq10}) is:} \quad p_1=\frac{2rGF''+F'(1G+rG')+FG'}{rF}\,,
\end{align}
where the components of the energy-momentum tensor $T^\mu_\nu$ are given by $T^\mu_\nu=[\epsilon,\, -p,\, -p_1,\, -p_1]$,
where $\epsilon$,  $p$, and $p_1$ are the density, radial and tangential  pressures, respectively.

The three differential equations (\ref{feqq}) are nonlinear and involve five unknowns: $F$, $G$, $\epsilon$, $p$, and $p_1$.  Thus, in order to make the aforementioned system solvable, two additional conditions are necessary. One possibility is to assume the time part of the metric potential $F$ which is shaped as \cite{Torres-Sanchez:2019wjv,Estevez-Delgado:2018bxa}:
\begin{equation}\label{metg}
F(r)={\frac {c_0 \left( 5+4\,c_1{r}^{2} \right) }{\sqrt {1+c_1{r}^{2}}}}\,,
\end{equation}
where $c_0$ is a dimensionless constant and $c_1$ is another constant with units of length denoted as $l^{-2}$.
Using the anisotropy equation involves utilizing the r\,r and $\mathrm{\theta\, \theta}$  of Eq. (\ref{feqq}), the second condition, and imposing Eq. (\ref{metg}) leads to the following result:
\begin{align}\label{metf}
&G(r)=  \frac{\left( 1+{c_1}{r}^{2} \right)^2}{\varphi^{3/2}}\biggl[(5+6\,c_1 r^2)\varphi- 4\,c_1(1+c_1{r}^{2}) {r}^{2}\varphi_1+c_2{r}^{2}(1+c_1{r}^{2}) \biggr] \,,
\end{align}
where $c_2$ represents an additional constant in terms of dimensions, $\varphi=\sqrt{{5+12\,c_1{r}^{2}+8\,{c_1}^{2}{r}^{4}}}$ and $\varphi_1={\arctanh} \left( {\frac {1+2\,c_1{r}^{ 2}}{\varphi}} \right)$. By utilizing equations (\ref{metg}) and (\ref{metf}) within the set of differential equations (\ref{feqq}), we obtain the energy density and pressure represented as:
\begin{align}\label{sol}
& \epsilon=\frac{1}{\varphi^{5/2}\kappa^2c^2  } \left[ 12c_1 \left( 8{c_1}^{3}{r}^{6}+16{c_1}^{2}{r}^{4}+15c_1{r}^{2}+5\right)  \left( 1+c_1{r}^{2} \right) ^{2}\varphi_1 -\left( 144{c_1}^{5}{r}^{8}+60c_1+486{c_1} ^{3}{r}^{4}+{c_1}^{2}{r}^{2}\right.\right.\nonumber\\
 & \left.\left.+408{c_1}^{4 }{r}^{6} \right)\varphi^{1/2}-3c_2 \left( 8{c_1}^{3}{r}^{6}+16{c_1}^{2}{r}^{4}+15c_1{r}^ {2}+5 \right)  \left( 1+c_1{r}^{2} \right) ^{2} \right]\,,\nonumber\\
& p=\frac{1}{\varphi^{3/2} \left( 5+4c_1{r}^{2} \right)\kappa^2} \left[ \left( 72{c_1}^{4}{r}^{6}+167{c_1}^{2}{r}^{2}+190{c_1}^{3}{r}^{4}+50c_1 \right) \varphi^{1/2}-4c_1 \left(12 {c_1}^{2}{r}^{4}+15c_1{r}^{2}+5\right)  \left( 1+c_1{r}^{2} \right) ^{2}\varphi_1\right.\nonumber\\
  & \left.+c_2\left(12 {c_1}^{2}{r}^{4}+15 c_1{r}^{2}+5\right)  \left( 1+c_1{r}^{2} \right) ^{2} \right]  \,.
\end{align}
Stressing that utilizing metric potentials (\ref{metg}) and (\ref{metf}) in equation (\ref{feqq}) leads to $p=p_1$, guaranteeing that our model possesses an isotropy. The mass within a sphere with a radius of $r$ is defined as:
\begin{align}\label{mas}
M(r)=4\pi {\int_0}^r \epsilon(\xi) \eta^2 d\xi\,.\end{align}

By plugging the energy-density form provided in Eq. (\ref{sol}) in Eq. (\ref{mas}), we obtain the mass's asymptotic form as:
\begin{align}\label{mas1}
& M(r)\approx
 - (0.63c_1+0.09c_2){r}^{3}+
 (0.18c_1+ 0.054c_2)c_1{r}^{5}- (0.38{c_1}+ 0.054c_2){c_1}^{2}{r}^{7}\,.\end{align}
The compactness parameter of a spherically symmetric object with radius $r$ is described as \cite{NewtonSingh:2019bbm,Roupas:2020mvs}:
\begin{align}\label{gm1}
& C(r)=\frac{2M(r)}{r}.
\end{align}
In the next section, we will explore the potential physical requirements for an isotropic stellar configuration and assess whether the model (\ref{sol}) satisfies these criteria.
\section{Essential requirements for a physically feasible isotropic model of a star}\label{S4}
Prior to moving forward, we will utilize the subsequent dimensionless replacement $r=xR.$
In this context, $R$ symbolizes the star's radius while $x$ symbolizes a parameter that has no units and has a value  one at the star's outer edge and zero at its core.  Furthermore, we expect that the dimensional variables $c_1$ and $c_2$ to be in  the form: $c_1=\frac{v}{R^2}\,, \qquad c_2=\frac{u}{R^2}$, with $v$ and $u$ being quantities with dimensions.   Plugging $c_1$, $c_2$ as well as $r$ into  Eqs. (\ref{metg}), (\ref{metf}) and (\ref{sol}), give non-dimensional parameters.   Now we list the essential requirements utilized in the isotropic model.

A model of isotropic star  should meet the requirements listed below in terms of the configurations of the star.
\begin{itemize}
\item  The components of the metric $F(x)$ and $G(x)$, as well as the energy-momentum tensor components $\epsilon$ and $p$, should exhibit favorable characteristics at the stellar object's center and display uniform behavior throughout the star's structure without any singularities.
\item Energy density, denoted by $\epsilon$, must remain nonnegative throughout the star's internal composition.  It must also have a finite positive quantity and gradually diminish as it approaches the star's interior.
\item It is a requirement for the pressure, denoted as $p$, to remain nonnegative within the fluid system. Also, the pressure gradient must be negative throughout the structure, represented as $\frac{dp}{dx}< 0$.
\item  The following inequalities are necessary for the isotropic star's energy conditions:
\begin{enumerate}[i.]
\item Within the confines of the weak energy condition (WEC), the inequality $p + \epsilon > 0$ holds.
\item The dominant energy conditions require that the $\epsilon$ is greater than or equal to the absolute value of $p$.
\item The Strong energy condition (SEC) states that $p$ plus $\epsilon$ must be positive, and $\epsilon$ plus 3$p$ must be positive.
\end{enumerate}
\item To ensure a practical model, the condition of causality should be confirmed, meaning the speed of sound $v$ in the star's interior structure must be less than 1, assuming the speed of light is 1.
\item The internal metric, $F$ and $G$, need to be seamlessly connected to the outer Schwarzschild metric at the boundary.
\item A stable star model requires the adiabatic index to be higher than $\frac{4}{3}$.
\end{itemize}
We are ready to evaluate our model against the physical criteria listed above to determine if it meets all of them.
\section{The actions of model (\ref{sol}) in the physical realm}\label{S5}
\subsection{Freedom of the model from singularity}
\begin{enumerate}[i.]
\item Equations (\ref{sol}) and (\ref{metf}) satisfy:
\begin{align}\label{sing}
F_{x\rightarrow 0}=5c_0\,\qquad  \qquad \textrm{and} \qquad \qquad {G}_{x\rightarrow 0}=1\,.
\end{align}
Equation (\ref{sing}) guarantees that $g_{00}$ and $g_{rr}$ are finite at the stellar core. Furthermore, the derivative of $F(x)$ and $G(x)$  with respect to $x$ should be limited at the center.  Equations (\ref{sing}) guarantee that $F(x)$ and $G(x)$  are smooth at the core and exhibit favorable characteristics throughout the star's core.
\item Equation (\ref{sol}) gives the density and pressure values at the star's core as:
\begin{align}\label{reg}
\epsilon_{_{_{x\rightarrow 0}}}= -0.12{\frac {15.7v+
 2.24u}{{R}^{2}{c}^{2}{\kappa}^{2}}}
\,,  \qquad \qquad {p}_{_{_{x\rightarrow 0}}}=0.04\,{\frac { 45.7v+ 2.24u}{{R}^{2}{\kappa}^{2}}}\,.
\end{align}
Equation (\ref{reg})  guarantees the non-negativity of $\epsilon$ and $p$ by assuming  \[- 15.7v-
 2.24u>0\,,\qquad   {\textrm and} \qquad  45.7v+ 2.24u>0\,.\]
Furthermore, the Zeldovich requirement \cite{1971reas.book.....Z} that links $\epsilon$ and $p$  at the star's as: $\frac{p(0)}{\epsilon(0)}\leq 1$  yields:
\begin{align}\label{reg1}
{\frac {- 20.4v- u}{ 21.1v+3u}}\leq 1 \Rightarrow u\geq-\frac{41.5}{4}v\,.\end{align}
\item The derivative of $\epsilon$ and $p$ given by Eq.  (\ref{sol}) yield the following form:
\begin{align}\label{dsol}
&\epsilon'=-\frac{8v x}{ {R}^{3}{\kappa}^{2}{c}^{2} \varphi^{4}} \left[ 12\varphi^{1/2}v Y  \left( 16{v}^{ 4}{x}^{8}+88{x}^{6}{v}^{3}+166{x}^{4}{v}^{ 2}+115v{x}^{2}+25 \right) \varphi_1-v \varphi \left(32 {v}^{4}{x}^{8}+96{x}^{6}{v}^{3}\right.\right.\nonumber\\
&\left.\left. +372{x}^{4}{v}^{2}+480v{x}^{2}+175 \right)-3 Y\left( 16{v}^{4}{x}^{8}+88{x}^{6}{v}^{3}+166{x}^{4}{v}^{2}+115v{x}^{2} +25 \right)c_1\varphi^{1/2}  \right]\,,\nonumber\\
& p'=-\frac{vx\left( 3+4v{x}^{2} \right) }{3{\kappa}^{2}{R}^{3} \left( 4v{x}^{2}+5 \right) ^{2} \varphi^{3}
 }\left[ 24 \left(24 {x}^{6}v^{3}+70{x}^{4}v^{2}+75v{x}^{2}+25\right) v\varphi^{1/2} \left( 1+v{x}^{2} \right) \varphi_1 -6 \left( 24{x}^{6}v^{3} \right.\right.\nonumber\\
&\left.\left.  +70{x}^{ 4}v^{2}+75v{x}^{2}+25\right)Y c_1\varphi-2v \varphi  \left(16 {x}^{6}v^{3}+14{x }^{4}v^{2}+40v{x}^{2}+25 \right) \right] \,, \,\mbox{with $Y=\left( 1+vx^2 \right)$.}
\end{align}
where $\epsilon'=\frac{d\epsilon}{dx}$ and  $p'=\frac{dp}{dx}$. Eqs.~(\ref{dsol}) show that the derivative $\epsilon$ and $p$  have negative values, as shown in Fig. \ref{Fig:2} \subref{fig:grad}.\\
\item The speed  of sound  gives:
\begin{align}\label{dso2}
&v^2=\frac{dp}{d\epsilon}=\frac{\varphi \left( 3+4vx^2 \right) c^2}{ 3\left(4 vx^2+5\right)^2 } \left[ 3 \left( 24x^6v^3+70x^4v^2+75vx^2+25 \right) Yv\varphi u-12 \left( 24x^6v^3+70x^4v^{2}+100vx^2+25 \right)\varphi^{1/2}\varphi_1 \right.\nonumber\\
&\left.+ 3v \varphi\left( 16x^6v^3+28x^4v^{2}+40vx^2+25 \right)   \right] \left[ \left(16 v^4x^8+11 {x}^{6}v^{3}+166x^4v^{2}+115v{x}^{2}+25 \right)  \left( 1+v{x} ^{2} \right) u\varphi -4v \left( 16 v^4{x}^8\right.\right.\nonumber\\
&\left.\left.+11x^6v^3+166x^4v^2+115vx^2+25 \right) \varphi^{1/2} \varphi_1+3v \varphi \left( 32v^4x^8+96x^6v^3+372x^4v^2+480vx^2175\right)  \right]^{-1}.
\end{align}
 Equation \eqref{dso2} is plotted in \ref{Fig:2}  \subref{fig:sped}
which demonstrates that sound travels at a speed lower than unity.
\end{enumerate}
\subsection{Junction requirements}
The exterior part of the star is thought to be empty and is described by the Schwarzschild metric, the only spherically symmetric solution in the symmetric teleparallel equivalent of general theory:
\begin{align}\label{Eq1}  ds^2= -\Big(1-\frac{c^2\kappa^2M}{4\pi xR}\Big)dt^2+\Big(1-\frac{c^2\kappa^2M}{4\pi xR}\Big)^{-1}dr^2+x^2R^2d\Omega^2,
 \end{align}
with $M$ being the overall mass of the system.
 We need to bring together the measurements of the interior (\ref{metg}) and (\ref{metf}) with the exterior Schwarzschild geometry (\ref{Eq1}) by joining them at the star's surface at  $x=1$. The point where the metric functions meet at the boundary $x=1$ gives rise to the following criteria:
\begin{align}\label{Eq2}  F(x\rightarrow1)=\Big(1-\frac{c^2\kappa^2M}{4\pi R}\Big), \qquad \qquad G(x\rightarrow1)=\Big(1-\frac{c^2\kappa^2M}{4\pi R}\Big)^{-1}\,.
 \end{align}
Moreover, we also set $c_0$, $v$ and $u$ of Eq. (\ref{metg}) and (\ref{metf}) in addition to enforcing pressure to vanish at the surface as:
\begin{align}\label{Eq3} & c_0=-{\frac {4\pi \,R\sqrt {1+v{R}^{2}}}{ \left( {c}^{2}{\kappa
}^{2}M-4\,\pi \,R \right)  \left(4 c_1{R}^{2}+5 \right) }}\,,\quad v={\frac {24\pi R-15{c}^{2}{\kappa}^{2}M\pm\sqrt {576{\pi }
^{2}{R}^{2}-80\pi R{c}^{2}{\kappa}^{2}M-15{c}^{4}{\kappa}^{4}{M}
^{2}}}{8 \left( 3{c}^{2}{\kappa}^{2}M-8\pi R \right) {R}^{2}}}\;,\nonumber\\
& u=\frac{ \left[ 16v{R}^{3}\pi  \left( 1+v{R}^{2} \right) ^{3}\varphi_1 + \left( 4\pi R-{c}^{2}{\kappa}^{2}M \right) \varphi^{3/2}-4 \left( 6v{ R}^{2}+5\right) \varphi \left( 1+v {R}^{2} \right) ^{2}R\pi  \right]}{4{\pi }{R}^{3} \left( 1+v{R}^{2 } \right) ^{3}}  \,.\nonumber\\
 \end{align}
 \section{Examination of solution (\ref{sol}) with viable compact stars}\label{S6}
We are now prepared to assess the model given by Eq.~(\ref{sol}) by applying the physical conditions mentioned earlier to estimate the observed pulsars masses and  radii. In order to obtain additional details about the model (\ref{sol}), we utilize the pulsar \textrm{HerX1} with  $M = 1.69\pm 0.15M_\circledcirc$ and $R =8.1\pm0.41$ km, correspondingly \cite{Gangopadhyay:2013gha}.
The mass is equal to $M=1.84 times the mass of the Sun,$ while the radius is $R=8.51 km$. The values of the dimensionless constants $c_0$, $c_1$, and $c_2$ are determined by these conditions:
\begin{align}\label{nv} c_0=0.2219294819\,,\qquad \qquad v=-15.53063284\,, \qquad u=1.136205369\,.\end{align}

 With the constants provided above, we graph the physical properties of the model (\ref{sol}).
\begin{figure}
\centering
\subfigure[~$\epsilon$ defined by Eq. (\ref{sol}) ]{\label{fig:dens}\includegraphics[scale=.23]{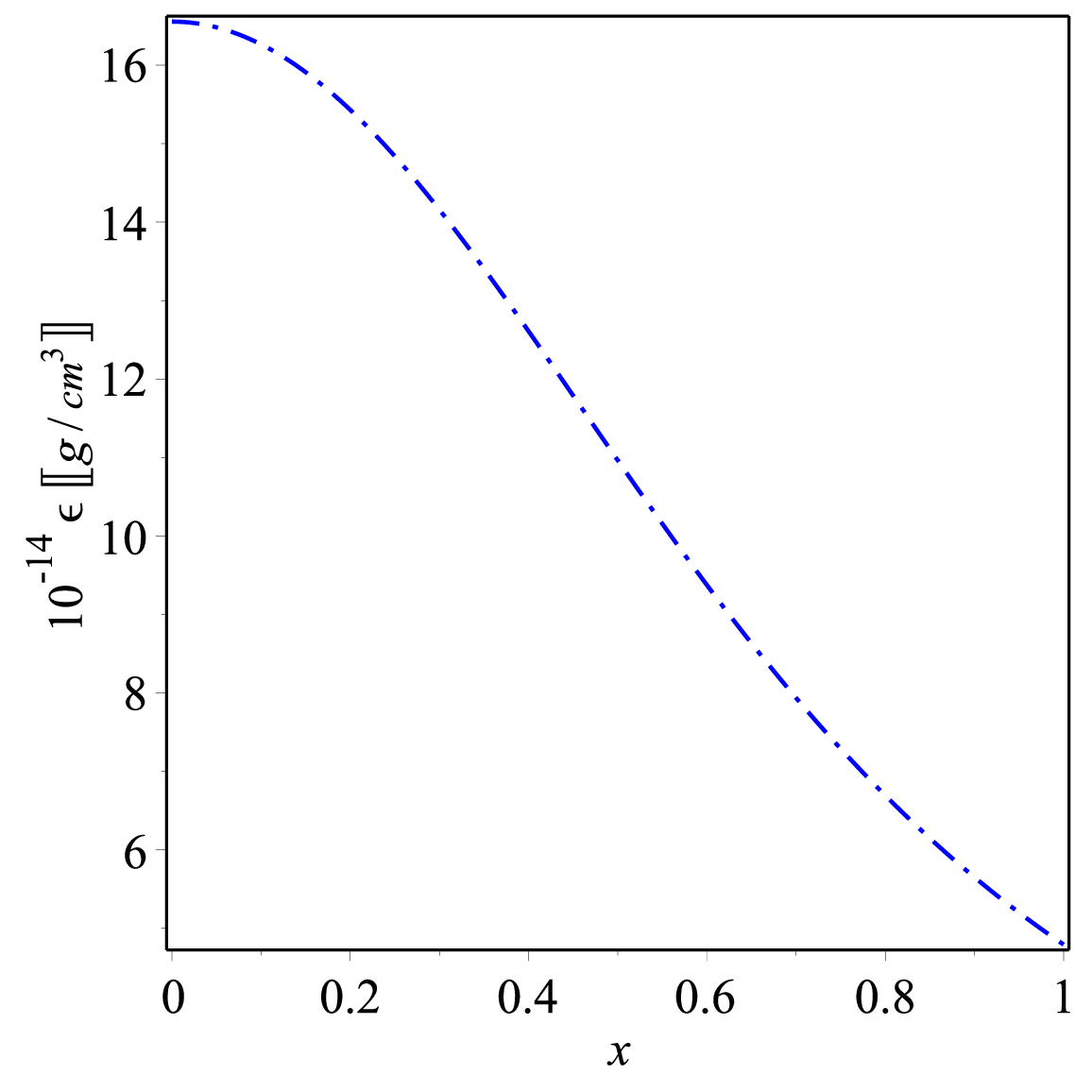}}
\subfigure[~ $p$ defined by Eq. (\ref{sol})]{\label{fig:pressure}\includegraphics[scale=.23]{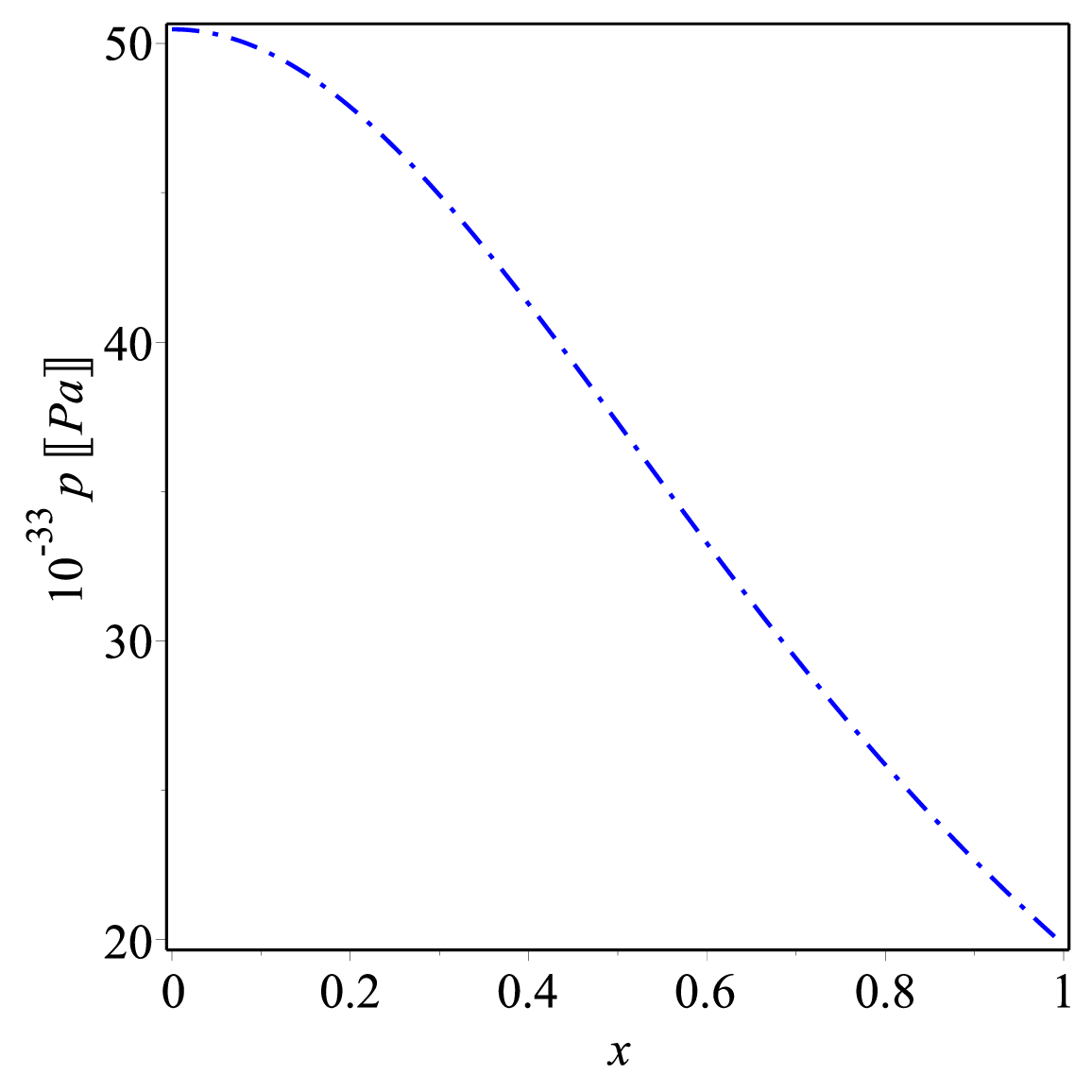}}
\caption[figtopcap]{\small{Figures \subref{fig:dens} and \subref{fig:pressure} display $\epsilon$ and $p$ of Eq.~(\ref{sol}) plotted against $x$, with constants fixed from pulsar \textrm{HerX1} \cite{Naik:2011qc}.}}
\label{Fig:1}
\end{figure}
Figures \ref{Fig:1} \subref{fig:dens} and \subref{fig:pressure} illustrate the energy-density behavior.

The density and pressure of the pulsar \textrm{HerX1} exhibit positive values required for proper stellar arrangement, with high density at the core and lower density towards the star's surface.
 Furthermore, figure \ref{Fig:1} \subref{fig:pressure} illustrates that at the stellar surface, the pressure is zero. The density and pressure patterns shown in figures \ref{Fig:1} \subref{fig:dens} and \subref{fig:pressure} are consistent with an accurate model.
\begin{figure}
\centering
\subfigure[~Derivatives of $\epsilon$ and $p$]{\label{fig:grad}\includegraphics[scale=0.23]{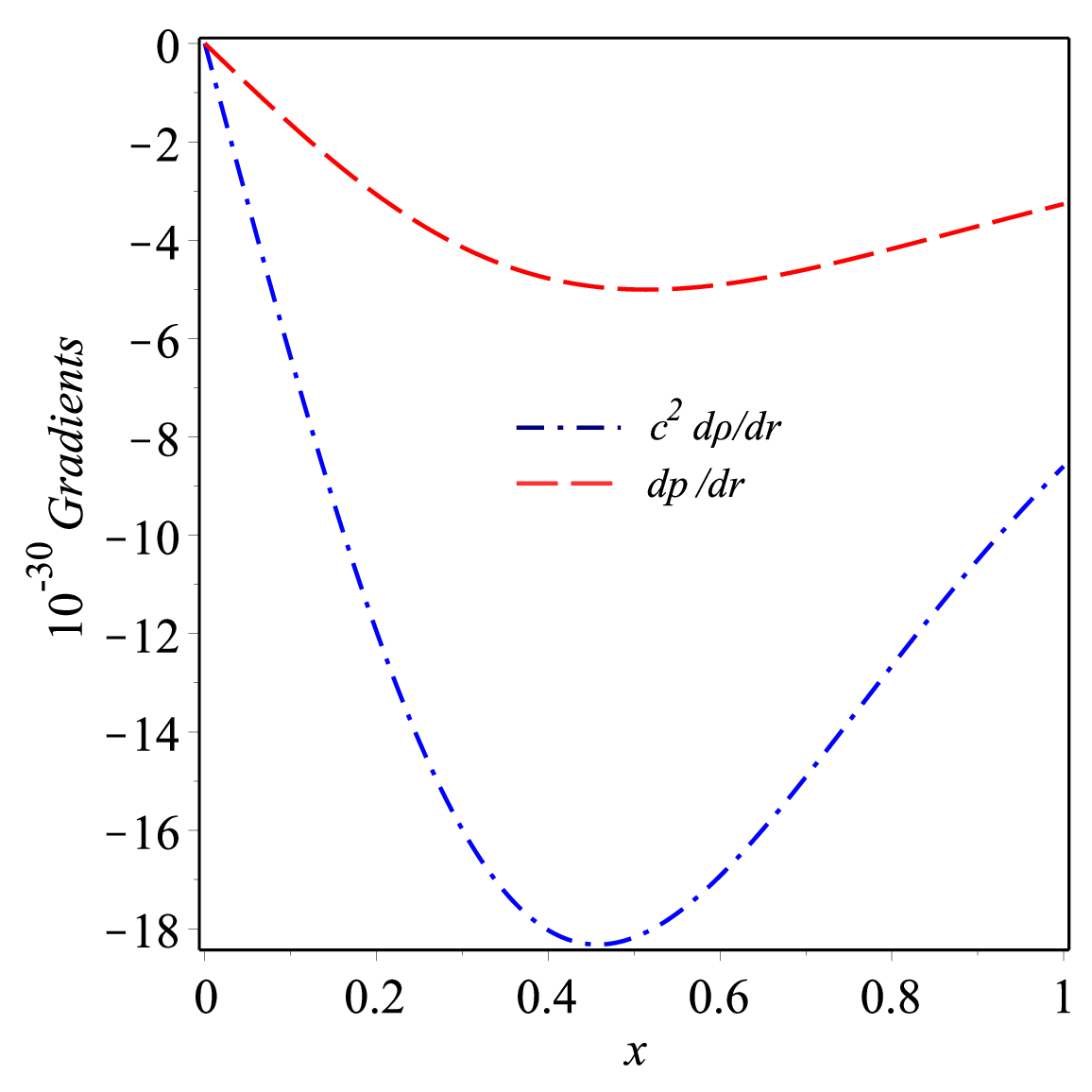}}
\subfigure[~Speed of sound]{\label{fig:sped}\includegraphics[scale=0.23]{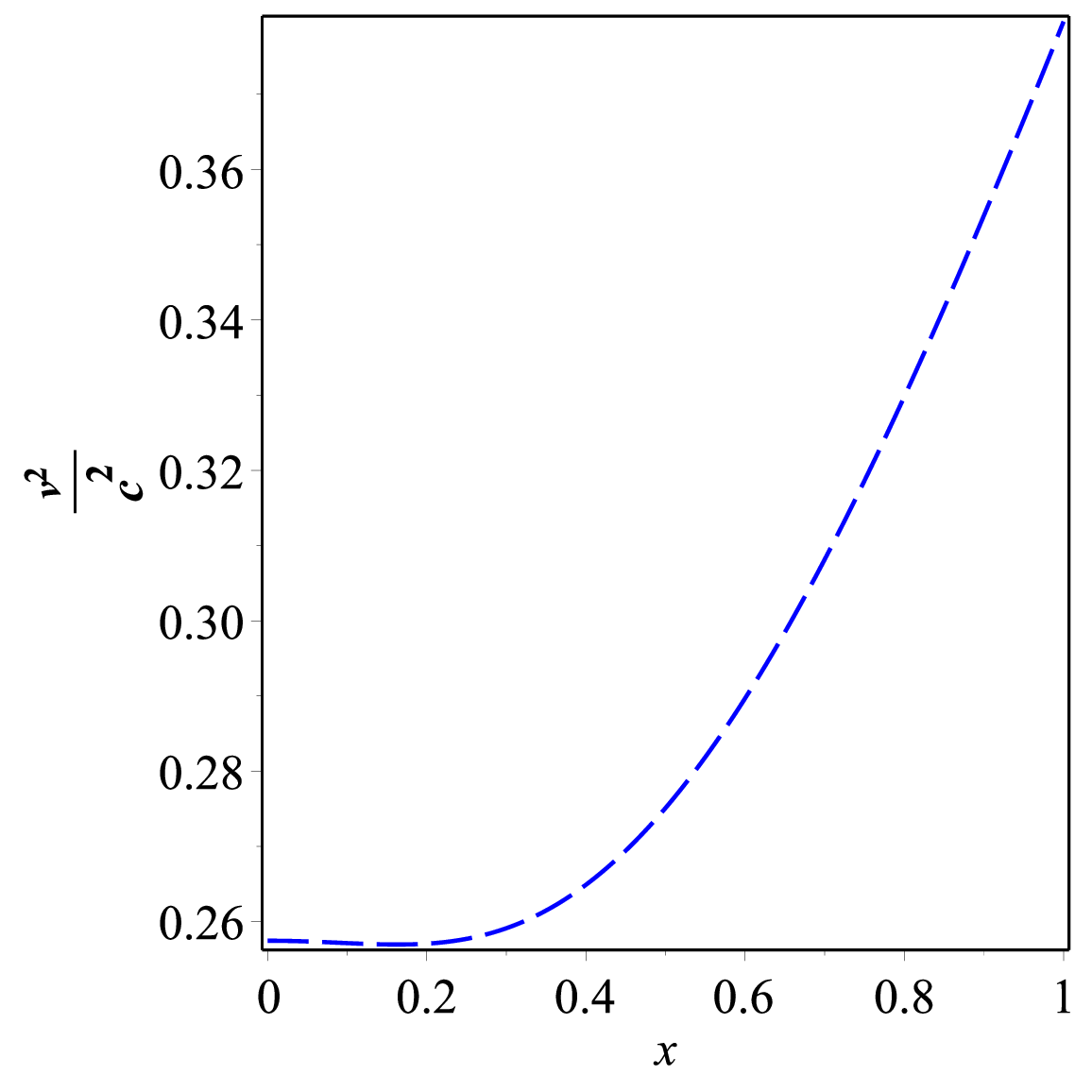}}
\subfigure[~Energy conditions]{\label{fig:EC}\includegraphics[scale=0.23]{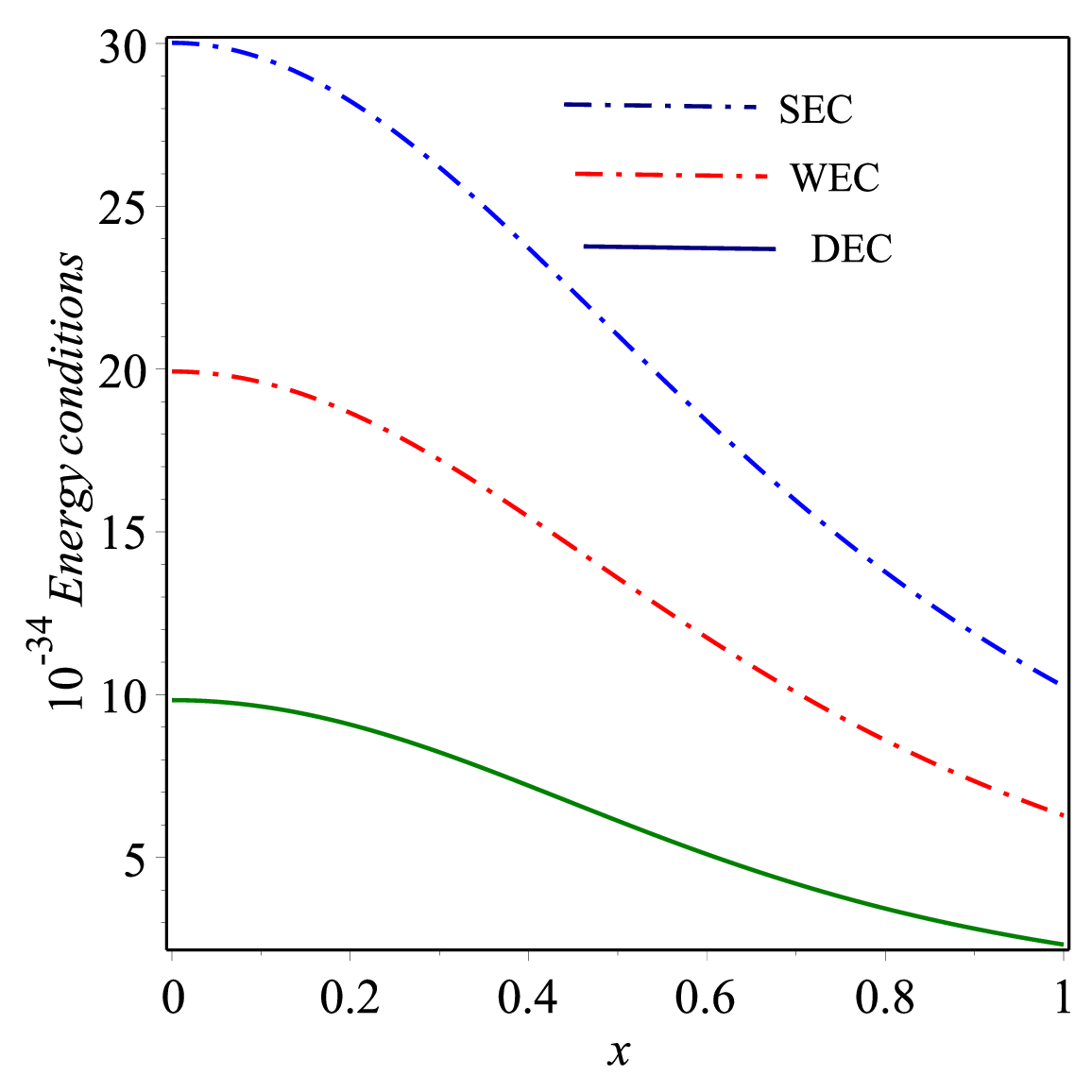}}
\caption[figtopcap]{\small{{{Plots: \subref{fig:grad} represents the derivatives $\epsilon$ and $p$; Fig. \subref{fig:sped} represents the  speed of sound, and \subref{fig:EC} symbolises the model's energy conditions through the dimensionless quantity $x$ with the constants fixed  by the pulsar  \textrm{HerX1}.}}}}
\label{Fig:2}
\end{figure}

 Fig. \ref{Fig:2} \subref{fig:grad}  demonstrates the negative nature of the pressure and density gradients, Fig. \ref{Fig:2}  \subref{fig:sped} confirming that the sound speed is below one, as needed for an accurate stellar model.
  Additionally, the energy conditions attitude is displayed in Fig. \ref{Fig:2}, \subref{fig:EC}. Therefore, the model configuration $\textrm{HerX1}$ meets all the requirements of the energy conditions for a significant and really isotropic stellar model.
\begin{figure}
\centering
\subfigure[~Equation of state $\omega=\frac{p(x)}{\epsilon(x)}$ ]{\label{fig:EoS}\includegraphics[scale=0.23]{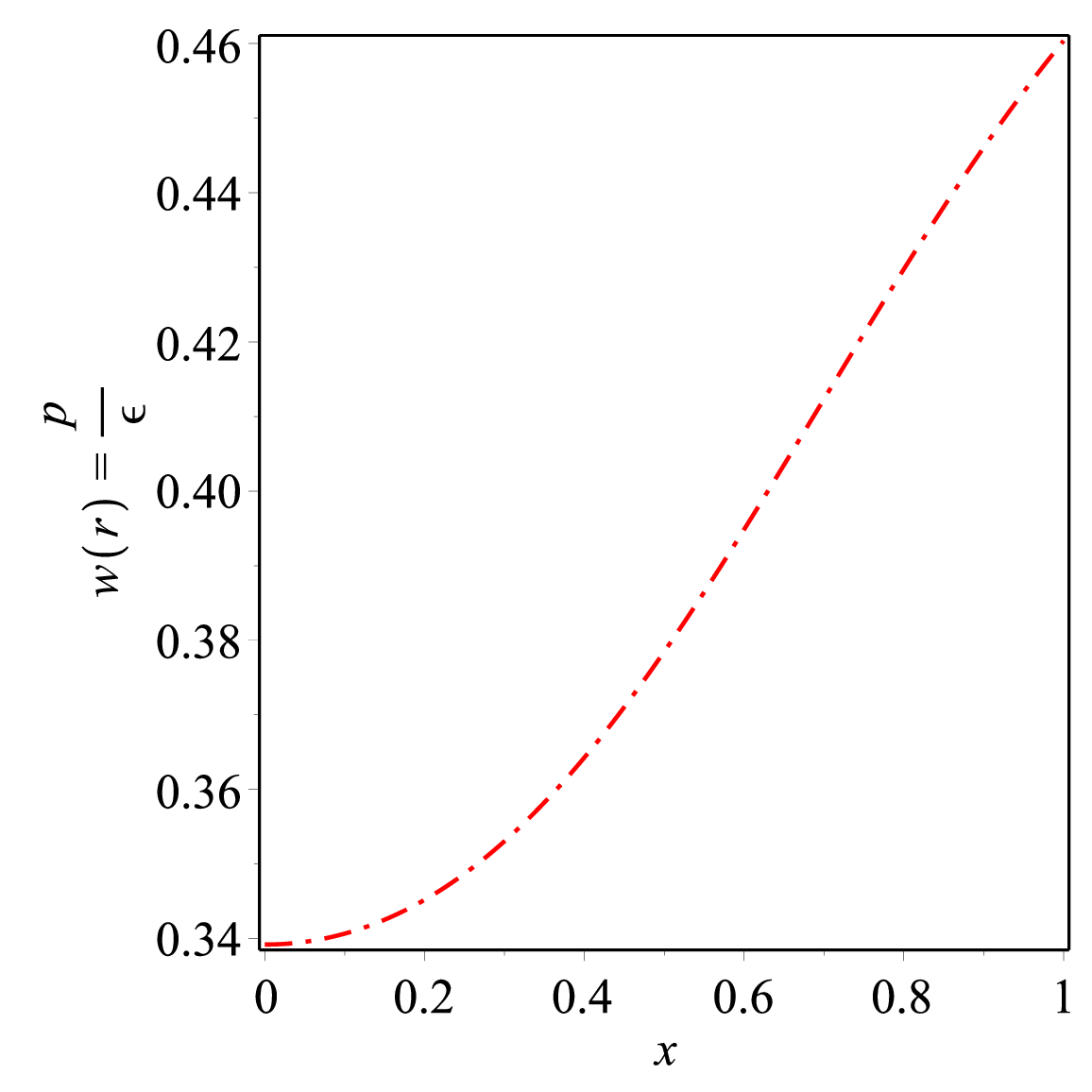}}
\subfigure[~The mass given by Eq.~(\ref{sol})]{\label{fig:mass}\includegraphics[scale=.23]{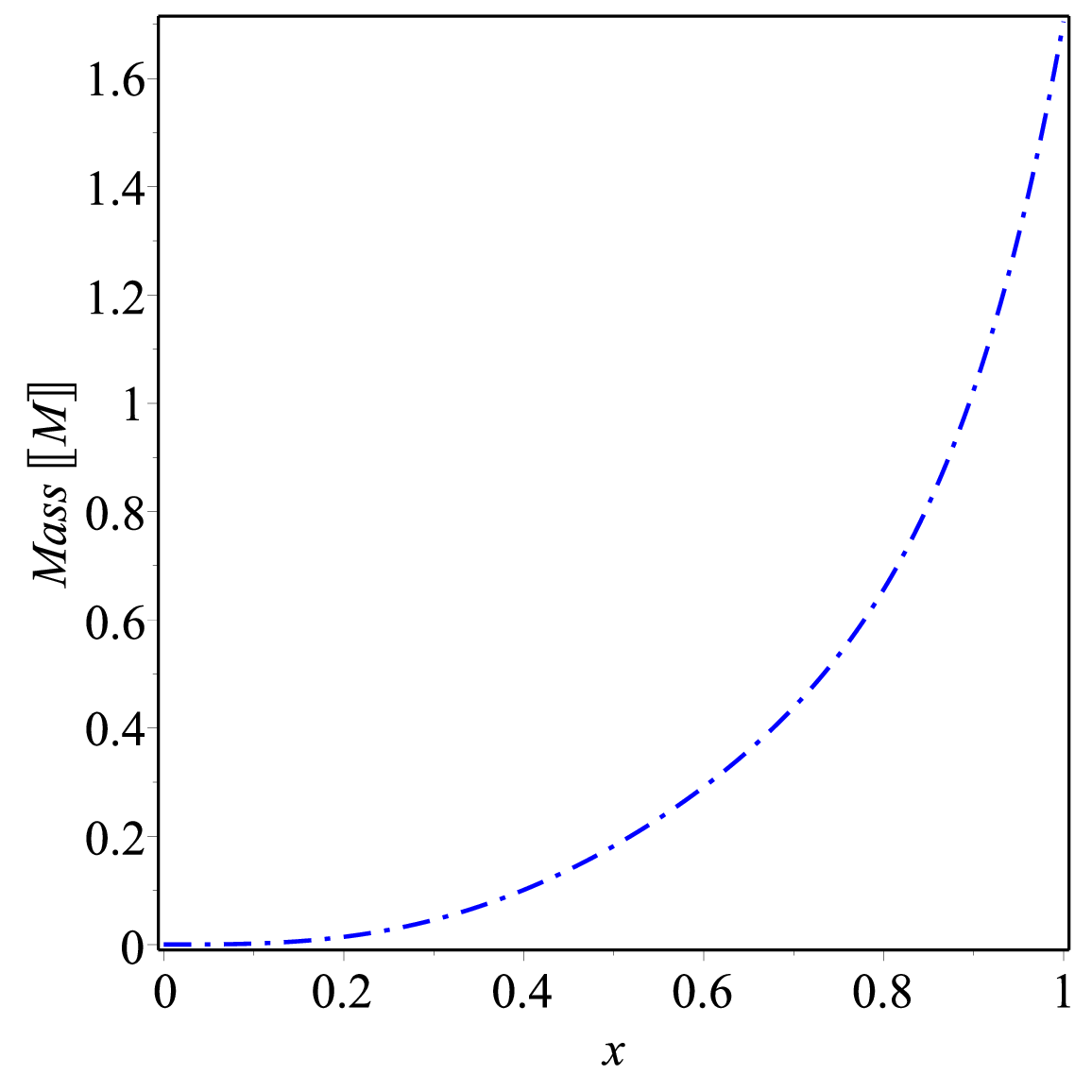}}
\subfigure[~$Z$ given by Eq.~(\ref{sol})]{\label{fig:red}\includegraphics[scale=.23]{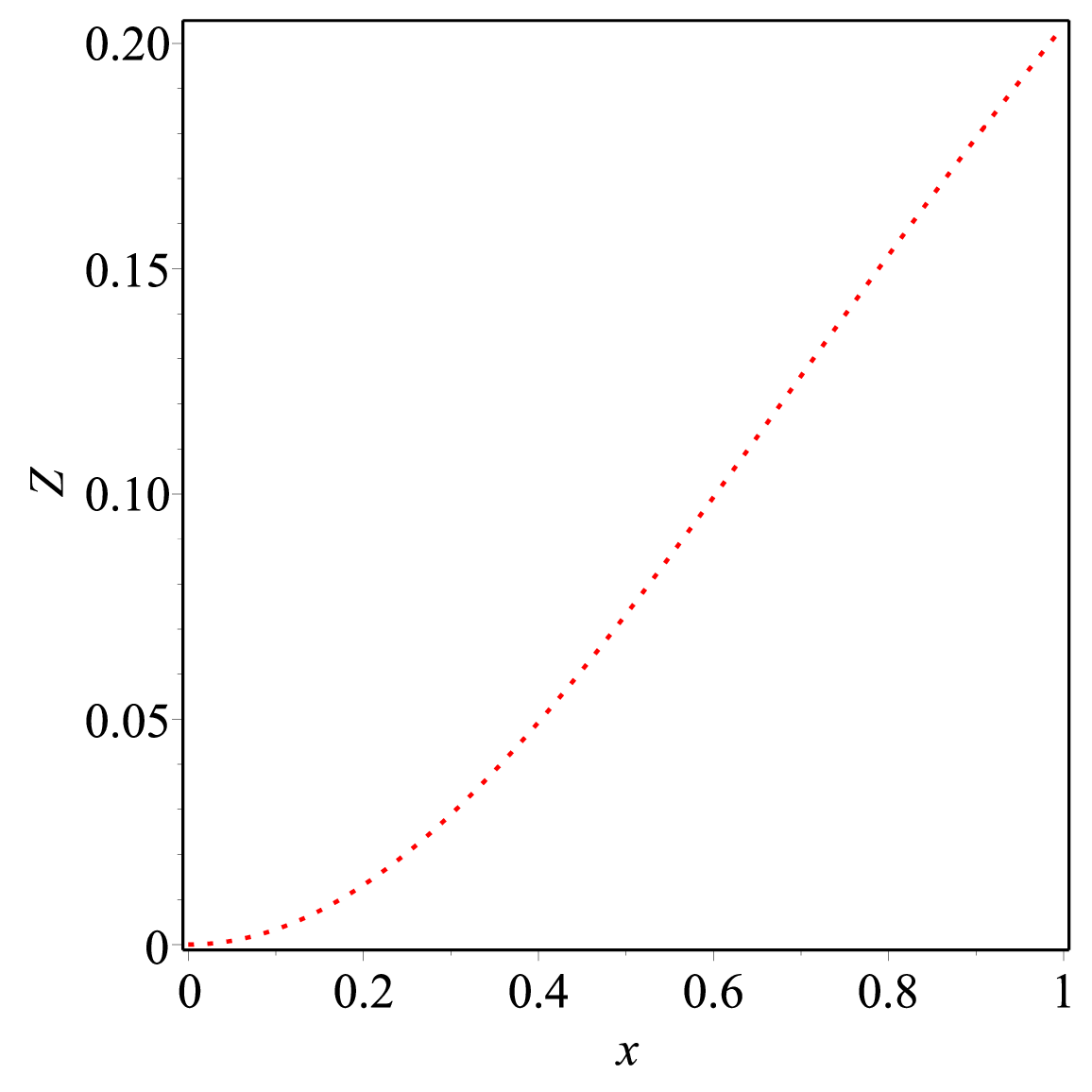}}
\caption[figtopcap]{\small{Graph of the equation of state (EoS) $\omega=\frac{p(x)}{\epsilon(x)}$ is plotted against the dimensionless variable x, utilizing the constants obtained from  \textrm{HerX1}, 
  \subref{fig:mass} illustrates how the mass  behaves, while \subref{fig:red} depicts the behavior of the red shift.}}
\label{Fig:3}
\end{figure}

Figure \ref{Fig:3} \subref{fig:EoS} illustrates the equation of state plotted against the dimensionless $x$, displaying a nonlinear pattern.

The mass function shown in Fig \ref{Fig:3} \subref{fig:mass} corresponds to the equation (\ref{mas}). Fig. \ref{Fig:3} \subref{fig:mass} shows the gradual increase in mass with the dimensionless $x$ and  shows that $M_{_{x=0}} = 0$. In the end, the trend in the red shift of the star is displayed in Fig. \ref{Fig:3} \subref{fig:red}. B\"{o}hmer and Harko \cite{Boehmer:2006ye} restricted the red-shift barrier to be equal to or less than $Z\leq 5$. The boundary redshift of the model, as calculated by $\textrm{HerX1}$, is $0.278269891$.

\section{Analysis of the stability of the model}\label{S7}
We will investigate the stability issue using two methods: the adiabatic index and the Tolman-Oppenheimer-Volkoff (TOV) equations.
\subsection{Assessing stability through the TOV equation}
Next, we consider the stability of equation (\ref{sol}) under the assumption of hydrostatic equilibrium.
 The TOV equation \cite{Tolman:1939jz,Oppenheimer:1939ne}  provides a formula for an isotropic:
\begin{align}\label{TOV}  {\mathrm -\frac{M_g(x)[\epsilon(x)+p(x)]{F}}{x\sqrt{G}}-\frac{dp}{dx}=0}\,,
 \end{align}
Here the gravitational mass $M_g(x)$ can be defined as:
\begin{align}\label{ma}  {\mathcal M_g(x)=4\pi{\int_0}^x\Big({T_t}^t-{T_r}^r-{T_\theta}^\theta-{T_\phi}^\phi\Big)\zeta^2 F\sqrt{G}d\zeta=\frac{xF'\sqrt{G}}{2F^2}}\,,
 \end{align}
Inserting Eq. (\ref{ma}) into (\ref{TOV}),  we get
\begin{align}\label{ma1}  -\frac{dp}{dx}{\mathrm -\frac{F'[\epsilon(x)+p(x)]}{2F}={\cal F}_g+{\cal F}_h=0}\,,
 \end{align}
 with ${\cal F}_g=-\frac{F'[\epsilon(x)+p(x)]}{2{F}}$ and ${\cal F}_h=-\frac{dp(x)}{dx}$ are, respectively, the gravitational and hydrostatic forces.

 The pulsar data from $\textrm{HerX1}$ is used to plot two distinct forces in Fig. \ref{Fig:4}. Fig. \ref{Fig:4} demonstrates that the pulsar experiences a balance between a positive gravitational force and a negative hydrostatic force, resulting in equilibrium.  Hence, we demonstrate the stability of the pulsar through the TOV equation by utilizing data from the pulsar $\textrm{HerX1}$.

\begin{figure}
\centering
\subfigure[~TOV ]{\label{fig:TOVgr}\includegraphics[scale=0.23]{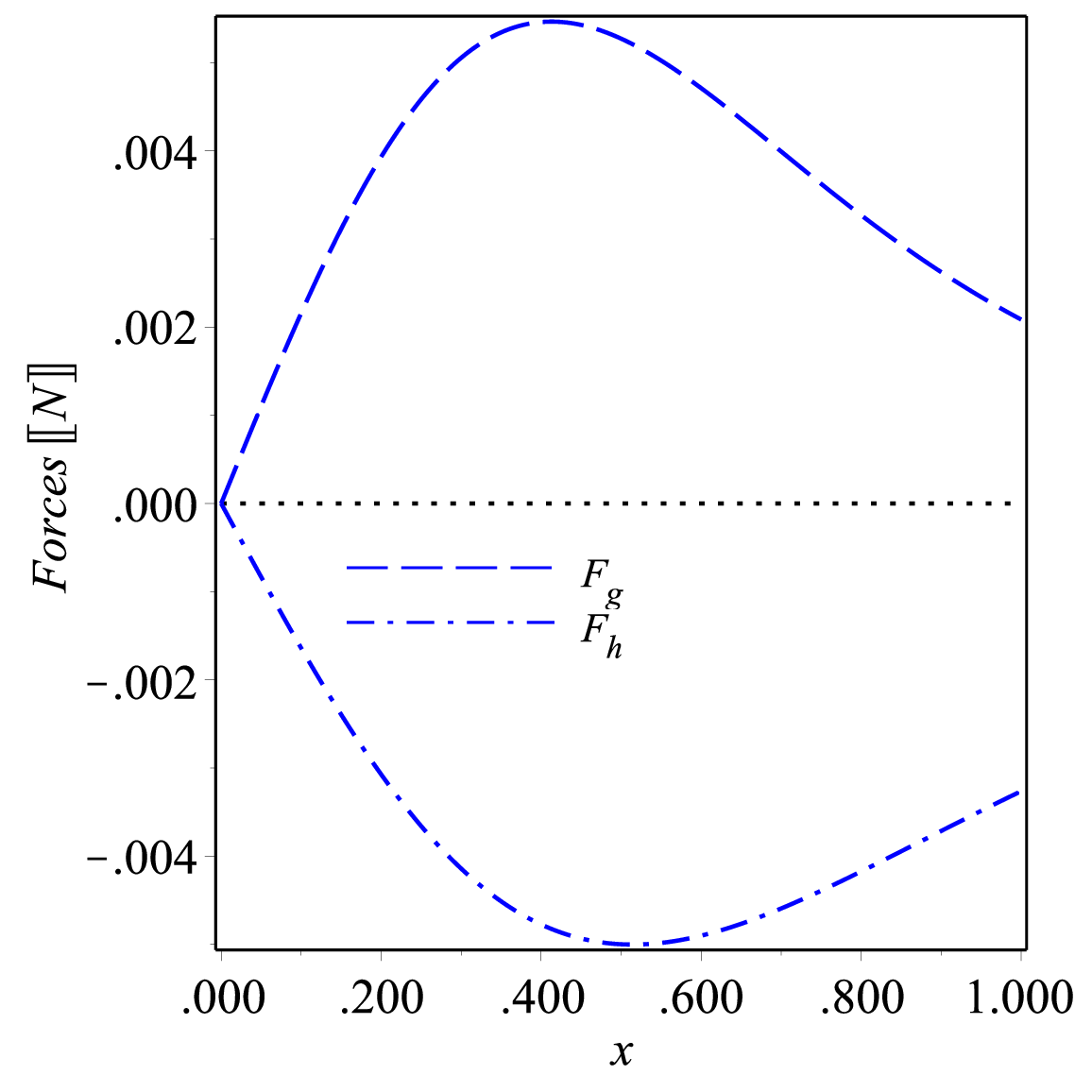}}
\subfigure[~$\Gamma$ ]{\label{fig:TOVras}\includegraphics[scale=.23]{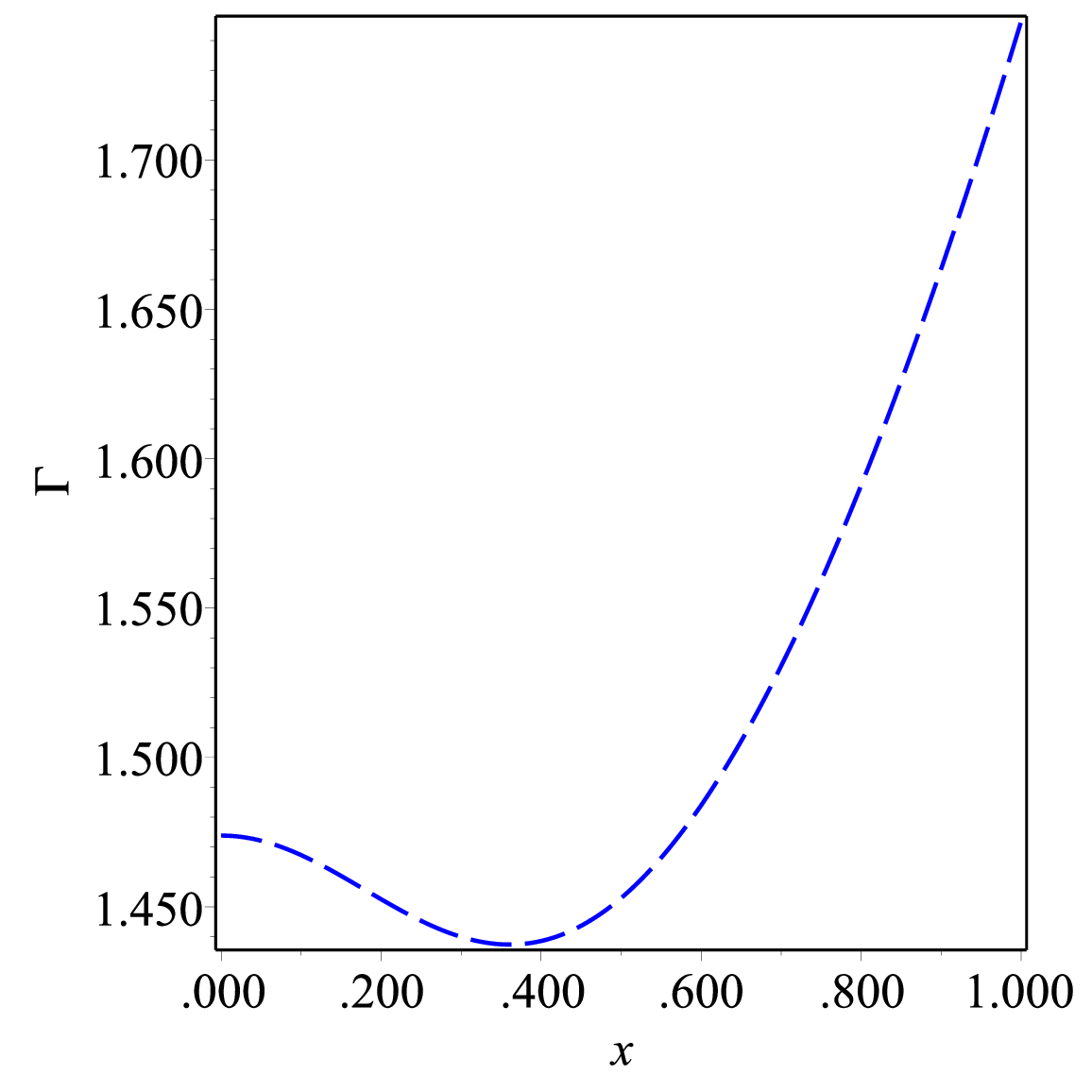}}
\caption[figtopcap]{\small{{Fig.~\subref{fig:TOVgr} illustrates the behavior TOV, while \subref{fig:TOVras} shows the adiabatic index. TOV and adiabatic index are plot against dimensionless $x$, where the dimensional constants are determined from observations of $\textrm{HerX1}$.}}}
\label{Fig:4}
\end{figure}
\subsection{Index of adiabaticity}
An alternative approach to assess the stability of the model involves analyzing the stability configuration using the adiabatic index, $\Gamma$, which serves as a key criterion. The adiabatic index is given by:
 \cite{Chandrasekhar:1964zz,1989A&A...221....4M,Chan:1993scw}
\begin{align}\label{a11}  {\mathrm \Gamma=\left(\frac{\epsilon(x)+p(x)}{p(x)}\right)\left(\frac{dp(x)}{d\epsilon(x)}\right)}\,.
 \end{align}
 In order for a Newtonian isotropic sphere to be in stable equilibrium, the adiabatic index must be considered $\Gamma$ should be $\Gamma>\frac{4}{3}$
 \cite{1975A&A....38...51H}.
An isotropic sphere has a neutral equilibrium when $\Gamma$ equals $\frac{4}{3}$.
 The adiabatic index of the model (\ref{sol}) can be calculated  and from this calculation we draw it in
 Fig. \ref{Fig:4} \subref{fig:TOVras}, we have illustrated $\Gamma$ indicating that its values exceed $4/3$ within the interior model, thereby meeting the stability requirement.
\section{ Relationship between mass and radius and equation of state}\label{Sec:EoS_MR}

Numerous astrophysicists are still trying to solve the mystery of the material makeup inside the cores of neutron stars.  This puzzle is generated due to the fact that the central densities of neutron stars rise to magnitudes multiple times higher than the nuclear saturation density, a domain outside the range of earthly laboratories.  Although we still don't fully understand the equation of state that determines how matter behaves in neutron stars, there is optimism that studying the mass and radius of neutron stars through astrophysical observations can provide useful limitations or rule out certain scenarios.  As a result, these astrophysical observations could be used to establish the mass-radius diagram linked to a specific equation of state.  Our study does not impose specific EoS; rather, we utilize the ansatz detailed in Eq. \eqref{metg} along with the anisotropy's disappearance.  By using the numerical data from Eq.~(\ref{nv}) for the pulsar $\textrm{HerX1}$ and applying the field equations of STEGR gravity, we create these sequences, illustrated in Fig. \ref{Fig:EoS1} \subref{Fig:EoS}.
\begin{figure}[th!]
\centering
\subfigure[~EoS ]{\label{Fig:EoS}\includegraphics[scale=0.25]{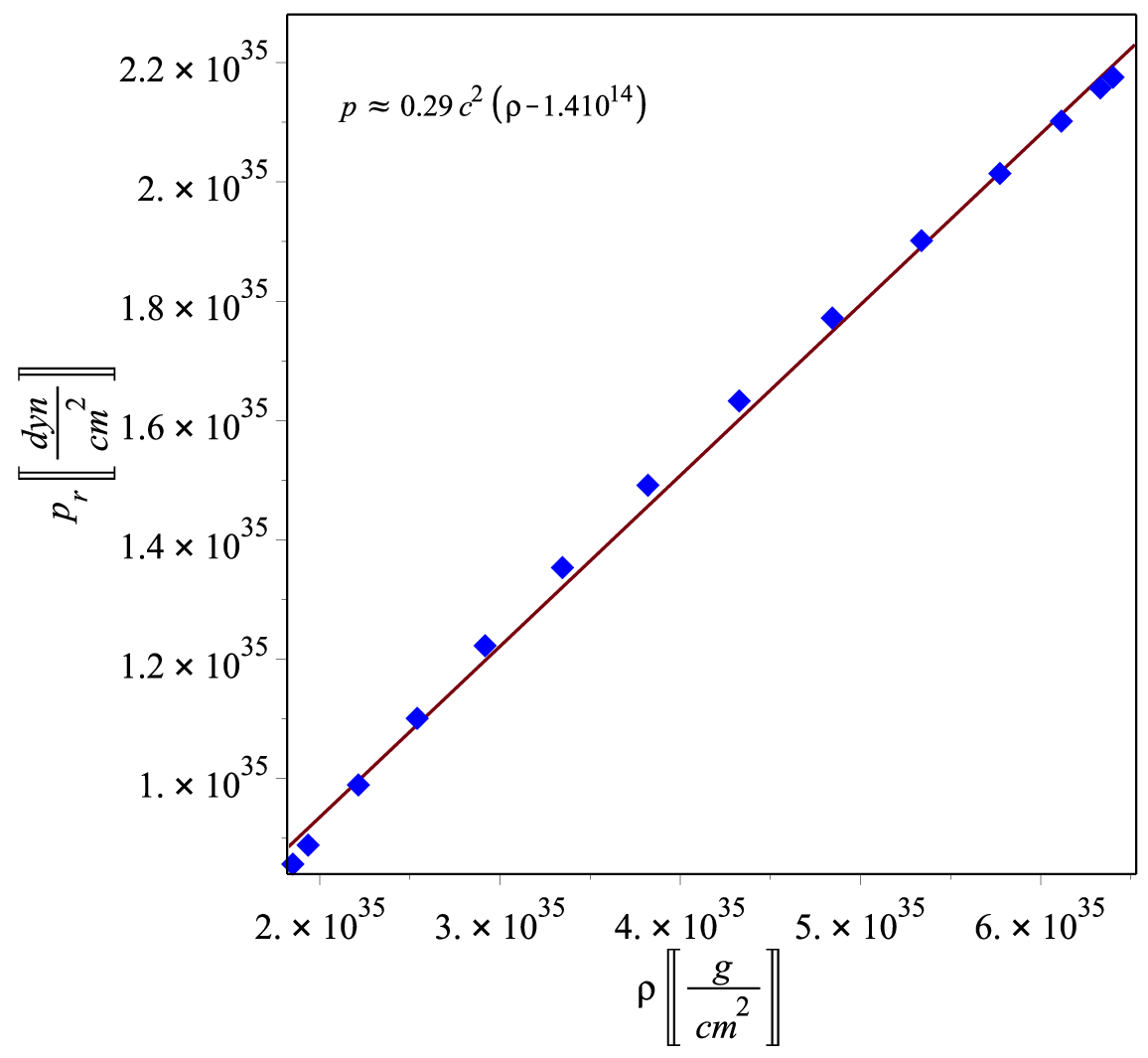}}
\subfigure[~Compactness-versus-radius plot]{\label{fig:Comp}\includegraphics[scale=0.25]{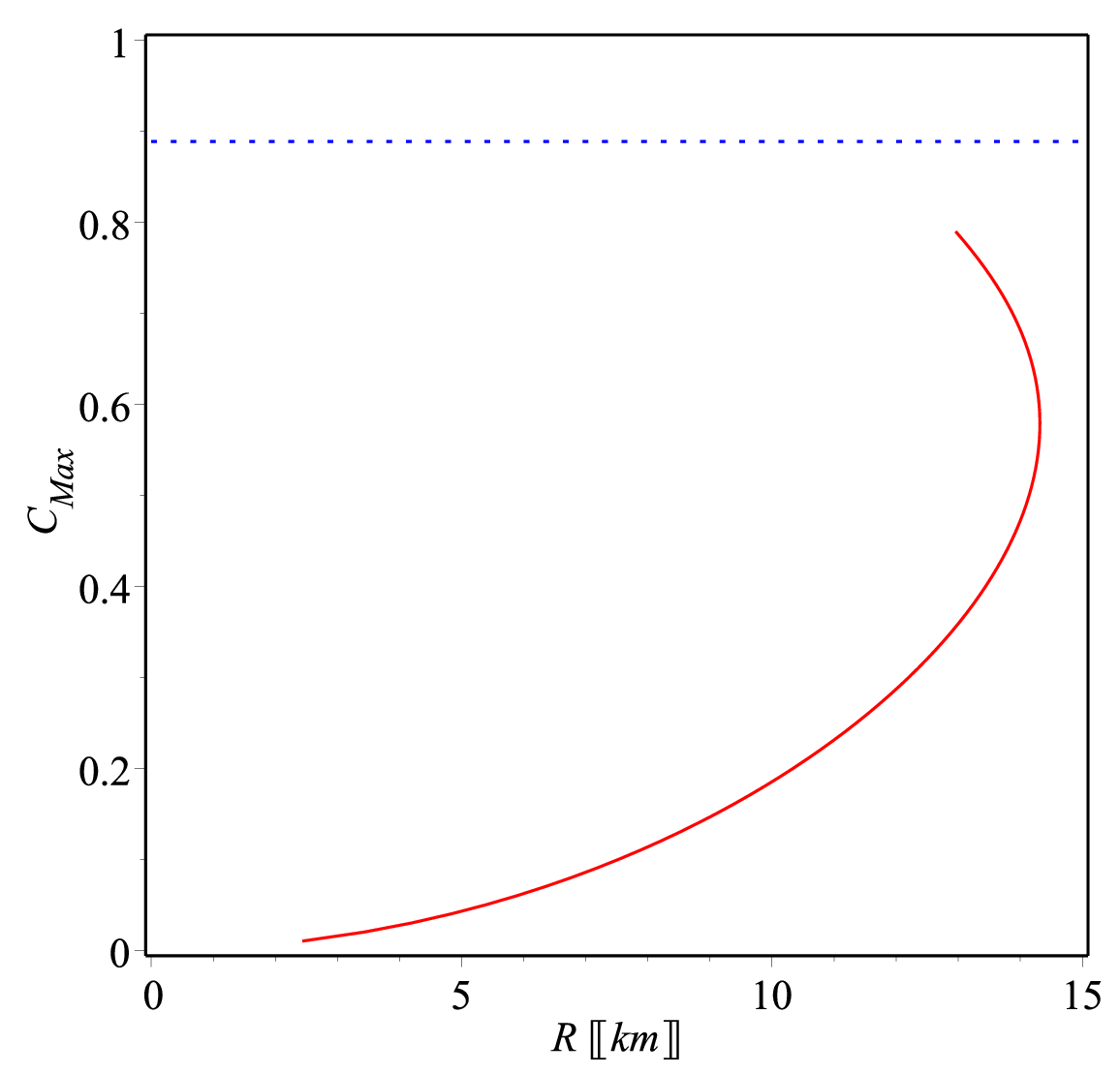}}
\subfigure[~Mass-versus-radius graph]{\label{fig:MR}\includegraphics[scale=.25]{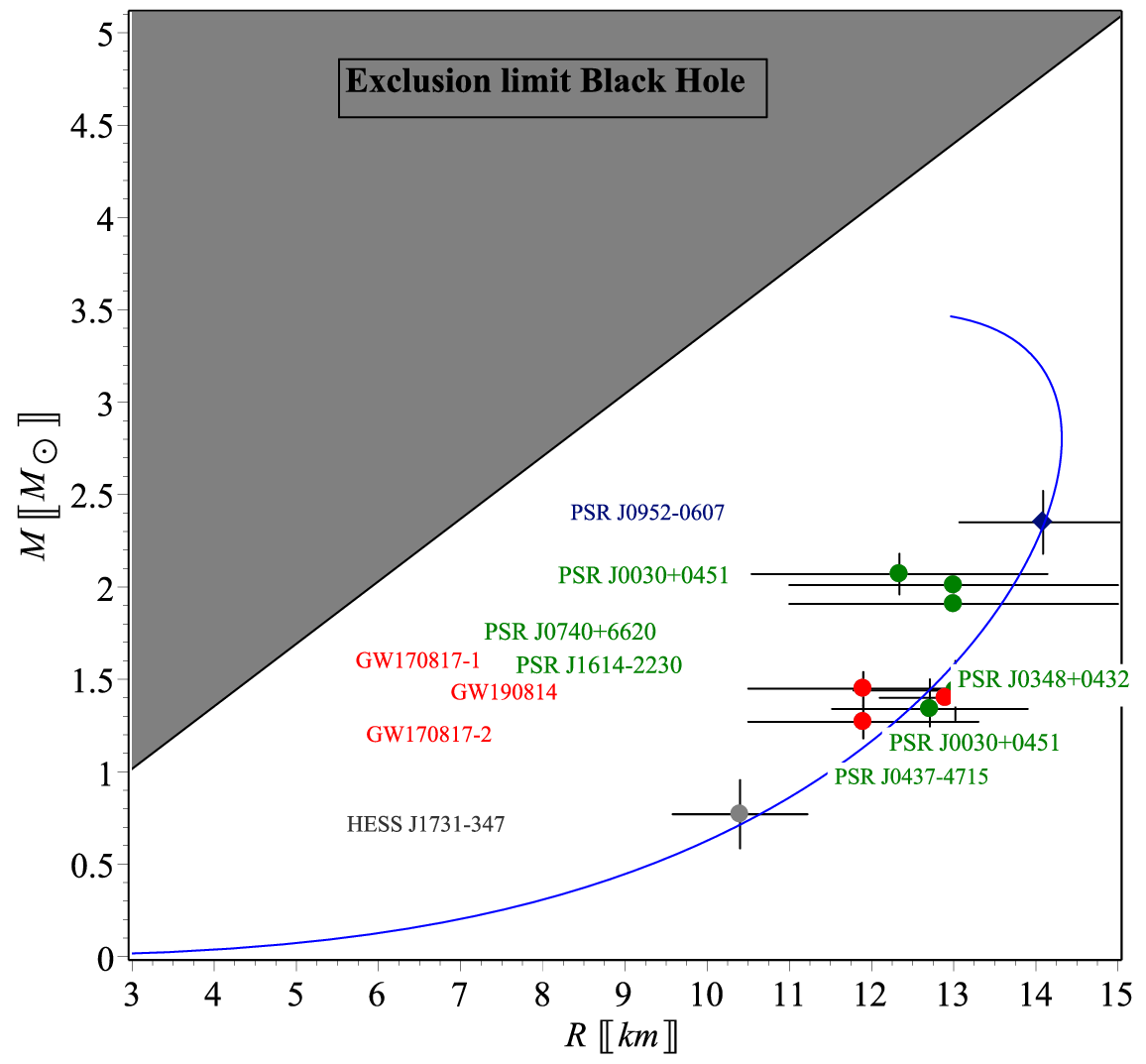}}
\caption{The data clearly indicates strong alignment with a linear model. The best equations can be shown as: $p \text{[dyn/cm$^2$]}\approx 0.29 c^2(\epsilon_s-1.4 \times 10^{14}\text{[g/cm$^3$]})$. \subref{fig:Comp}: The compactness-radius curves indicates that the maximum compactness is $C= 0.81$. \subref{fig:MR} The mass-radius curves show  an upper mass of $M\approx 3.47 M_\odot$ with radius ${R_s}\approx 13.42$ km.}
\label{Fig:EoS1}
\end{figure}
Clearly, the data shows a high level of agreement with a linear model. The most suitable equations can be represented as: $p \text{[dyn/cm$^2$]}\approx 0.29 c^2(\epsilon_s-1.4 \times 10^{14}\text{[g/cm$^3$]})$.

Buchdahl set a critical limit on stable stellar configurations, stating that the compactness value must be less than 8/9, as mentioned in \cite{PhysRev.116.1027}. Significantly, this limit was first established for isotropic spherically symmetric solutions in GR, which our model matched.

For the pulsar {\textrm{HerX1}}, let us calculate the Buchdahl limit:
Firstly, we take the following  value, $\epsilon_s=1.4\times 10^{14}$ g/cm$^3$ and solve the density profile given by Eq.(\ref{sol}) for the radius $R$ at the surface achieved from the best-fit EoSs, for any values of the compactness parameter $0 \leq C \leq 1$. Despite reaching a maximum compactness of $0.81$, similar to GR was observed by Roupas et al. (2020).

In this section, we provide the Mass-Radius curves corresponding to the best fit EoS as previously obtained, shown in Fig. \ref{Fig:EoS1}\subref{fig:MR}. Hence, we select the density boundary as $\epsilon_I=1.4\times 10^{14}$ g/cm$^{3}$ resulting in a maximum mass $M\approx 3.47 M_\odot$ at a radius of $R\approx 13.42$ km.

\begin{table*}[t!]
\caption{\label{Table1}%
The numerical values of the model parameters using different pulsars}
\begin{ruledtabular}
\begin{tabular*}{\textwidth}{lccccccc}
{{Pulsar}}                 &Ref.            & Mass ($M_{\odot}$) &        radius observation [{km}]   &    {$c_0$}    & {$v$} & {$u$}&\\ \hline
Her X-1     &    \cite{Abubekerov:2008inw}        &  $0.85\pm 0.15$    &      $8.1\pm0.41$ &     $0.2219294819$         &$-15.53063284$ & $1.136205369$     \\
Cen X-3 &\cite{Naik:2011qc}         &  $1.49\pm 0.49$    &  $9.178\pm 0.13$ &$0.2840689057$&  $-20.65405657$    & $1.333282132$\\
4U1608 - 52  &\cite{1996IAUC.6331....1M}              &  $1.57\pm 0.3$     &  $9.8\pm 1.8$  &     $0.2835722413$        &$-19.82321351$    &$1.274105446$  \\
EXO 1745-268  &\cite{Ozel:2008kb}   &  $1.65\pm 0.25$    &  $10.5\pm 1.8$   &     $0.2827131224$      & $-18.72033089$     &$1.19582314$ \\
4U 1820-30 &\cite{Guver:2010td}      &  $1.46\pm 0.2$     &  $11.1\pm 1.8$   &    $0.2711612677$ &$-12.13950334$        &$0.7378016429$\\
\end{tabular*}
\end{ruledtabular}
\end{table*}
\begin{table*}[t!]
\caption{\label{Table1}%
Values of physical quantities}
\begin{ruledtabular}
\begin{tabular*}{\textwidth}{lcccccccccc}
{{Pulsar}}                              &{$\epsilon\lvert_{_{_{x\rightarrow0}}}$}{[$g/cm^3$]} &      {$\epsilon\lvert_{_{_{x\rightarrow 1}}}$}{[$g/cm^3$]}  &   {$\frac{dp}{d\epsilon}\lvert_{_{_{x\rightarrow0}}}$}  &    {$\frac{dp}{d\epsilon}\lvert_{_{_{x\rightarrow 1}}}$}&
 {$(\epsilon-3p)\lvert_{_{_{x\rightarrow 0}}}$}{[$Pa$]}&{$(\epsilon-3p)\lvert_{_{_{x\rightarrow 1}}}$}{[$Pa$]} &{$z\lvert_{_{_{x \rightarrow 1}}}$}&\\
\hline\\
Her X-1                        &$\thickapprox$4.5   &$\thickapprox$16.62    &$\thickapprox$0.26 & $\thickapprox$0.39& $\thickapprox$9.7& $\thickapprox$30& 0.2  \\
Cen X-3           &$\thickapprox$7.1& $\thickapprox$19.3 &$\thickapprox$0.246 & $\thickapprox$0.41& $\thickapprox$7.7& $\thickapprox$27.1& 0.31  \\

4U1608 - 52             &$\thickapprox$6.8& $\thickapprox$16.23&$\thickapprox$0.249& $\thickapprox$0.395& $\thickapprox$7.83& $\thickapprox$23.05& 0.26   \\
EXO 1785 - 248   &$\thickapprox$6.93& $\thickapprox$13.9&$\thickapprox$0.249 & $\thickapprox$0.387& $\thickapprox$7.78& $\thickapprox$18.78& 0.233\\
4U1820 - 30     &$\thickapprox$5.05& $\thickapprox$18.2&$\thickapprox$0.247 & $\thickapprox$0.327& $\thickapprox$5.4& $\thickapprox$10.38& 0.156
\end{tabular*}
\end{ruledtabular}
\end{table*}
A comparable analysis can also be applied to pulsars other than $\textrm{HerX1}$. In Tables I and II, we provide concise findings for the remaining observed pulsars.

\section{Discussions and Conclusion}\label{Conc}

Through the present study, we delved into the intriguing domain of isotropic stars within the framework of symmetric teleparallel theory. Our findings significantly enhance the existing body of knowledge in this field, particularly in understanding the behavior of these stars under the principles of symmetric teleparallel theory.

Our research demonstrates that the STEGR theory offers a robust framework for exploring the properties of isotropic stellar structures. This theory, which is characterized by a torsion-free connection, a zero Riemann tensor, and a nonzero nonmetricity scalar $Q$, has proven to be a powerful tool in our study. Within the context of STEGR gravitational theory, we have derived an isotropic solution without imposing any assumptions on the structure of the equation of state. The isotropic model is constructed by assuming the metric potential includes a temporal component and that anisotropy vanishes. One of the main features of this model is that it involves three dimensionless constants These constants are determined by matching the solution with the external Schwarzschild solution and ensuring that the pressure at the star's surface is zero. The following is a summary of the physical tests applied:

\begin{itemize}
\item The density and pressure within the stellar center  configuration are finite, while the pressure at the star's surface vanishes, as shown in Figs. \ref{Fig:1}  \subref{fig:dens} and \ref{Fig:1} \subref{fig:pressure}.
\item The observed downward trends in pressure and density, as shown in Fig. \ref{Fig:2} \subref{fig:grad}, the affirmation of causality demonstrated in Figs. \ref{Fig:2} \subref{fig:sped}, and the successful verification of energy conditions illustrated in Fig. \ref{Fig:2} \subref{fig:EC}, all contribute to a comprehensive and professional analysis of the system.
 \item Furthermore, we have demonstrated that the EoS parameter exhibits non-linear behavior, a distinct characteristic of the isotropic model, as shown in Fig. \ref{Fig:3} \subref{fig:EoS}. Additionally, we provided evidence of increasing mass, as illustrated in Fig. \ref{Fig:3} \subref{fig:mass}, and observed that the model's surface exhibits a redshift value of
$Z=0.2782$, as depicted in Fig. \ref{Fig:3} \subref{fig:red}.
\item One of the advantage of this model is its successful validation of the TOV equation, as illustrated in Fig. \ref{Fig:4} \subref{fig:TOVgr}. It further offers an exact calculation of the adiabatic index, as presented in  Fig. \ref{Fig:4} \subref{fig:TOVras}.
\item Moreover, we demonstrate that the influence of isotropy could potentially lead to a pulsar mass exceeding the predictions of  GR  \cite{Astashenok:2021btj}, as illustrated in Fig. \ref{Fig:EoS1} \subref{fig:MR}.
\end{itemize}

Furthermore, we examined our model in the context of six additional pulsars, which enabled the numerical derivation of the constant values. Numerical calculations were performed for the density, equation of state parameter, both the strong energy condition and the redshift measured at the center and surface of the star. All these results are systematically presented in Tables I and II.

  In this study, we assumed one form of the ansatz given by Eq.~\eqref{metg} and used it in the anisotropic field equations to derive the other form of the metric ansatz. However, modified gravitational theories such as $f(T)$, $f(R)$, and $f(Q)$ do not permit the application of this procedure. This is because the anisotropic field equations in these theories are significantly more complex, making it challenging to derive other components of the metric ansatz. The difficulty primarily arises from the contributions of higher-order fields in these theories such as higher-order torsion scalars in $f(T)$, higher-order Ricci scalars in $f(R)$, and non-metricity contributions in $f(Q)$.

To derive isotropic spherically symmetric solutions in modified gravitational theories, one must either assume both ansatz forms of the metric that nullify the anisotropic field equations or develop an alternative approach. This task will be addressed in our future work.}



\end{document}